\documentclass[
               ]{jetp} 

\newcommand{\be}{\begin{equation}}
\newcommand{\ee}{\end{equation}} 
\newcommand{\bea}{\begin{eqnarray}}
\newcommand{\eea}{\end{eqnarray}}

\begin{document}
\English

\title{Stochastic bistable systems, and competing hysteresis and phase coexistence}




\author{Mahendra~K.}{Verma} 
\email{mkv@iitk.ac.in}
\affiliation {Department of Physics, Indian Institute of Technology Kanpur, Kanpur 208016, India}
\author{Abhishek}{Kumar}
\email{abhishek.kir@gmail.com}
\affiliation{Applied Mathematics Research Centre, Coventry University, Coventry CV15FB, The United Kingdom}
\author{Adhip}{Pattanayak}
\affiliation {Department of Physics, Indian Institute of Technology Bombay, Powai 400076, India}

\abstract{In this paper we describe the solution of a stochastic bistable system from a dynamical perspective.  We show how a single framework with variable noise can explain hysteresis at zero temperature and two-state coexistence in the presence of noise. This feature is similar to the phase transition of thermodynamics.  Our mathematical model for bistable systems  also explains how the width of a hysteresis loop shrinks in the presence of noise, and how variation in initial conditions can take such systems to different final states.}

\maketitle


\section{Introduction}

Many systems in nature make transition from one state to another depending on the parameter values.  For example, in a magnetic system, a paramagnetic state transforms to a ferromagnetic state depending on the temperature of the heat bath~\cite{Landau:book:StatMech,Reichl:book:StatMech,Pathria:book,Schwabl:book,Chaikin:Book}.  Such thermodynamic transition, to a mean field approximation, is described by the free energy of the form
\be
F = d(T) X^2 + X^4,
\label{eq:F_phi4}
\ee
where $X$ stands for the order parameter, here magnetization, and $d(T) \propto T-T_c$ with $T,T_c$ as the temperature and critical temperature respectively~\cite{Landau:book:StatMech,Reichl:book:StatMech,Pathria:book,Schwabl:book,Chaikin:Book}.  It has been shown that the aforementioned  transition and the liquid-vapor transition near the critical temperature belong to the same class, known as {\em second order transition} since the order parameter grows smoothly from zero to nonzero value~\cite{Landau:book:StatMech,Wilson:PR1974}.  However, far away from $T_c$, these systems exhibit first order transition in which the order parameter jumps by a finite value. To a mean field approximation, the free energies of such systems are approximated by 
\be
F = d(T) X^2 - X^4  + X^6.
\label{eq:F_phi6}
\ee
The aforementioned transitions are generally studied in the framework of equilibrium statistical mechanics~~\cite{Landau:book:StatMech,Reichl:book:StatMech,Pathria:book,Schwabl:book,Chaikin:Book}.

Interestingly, dynamical systems too exhibit similar transitions, which are often studied in the framework of {\em bifurcation theory}~\cite{Strogatz:book}.  Corresponding to Eqs.~(\ref{eq:F_phi4}, \ref{eq:F_phi6}),  the time-dependent variation of the order parameter are described by the following equations:
\bea
\dot{X}  &= &  cX - X^3,  \label{eq:X3} \\ 
\dot{X} &  = &  c X + X^3 - X^5.  \label{eq:X5}
\eea
Equation~(\ref{eq:X3}) exhibit supercritical  pitchfork bifurcation, while  Eq.~(\ref{eq:X5}) exhibits subcritical  pitchfork and saddle-node bifurcations, leading to hysteresis~\cite{Strogatz:book}. The models of Eqs.~(\ref{eq:X3}, \ref{eq:X5}) have been employed to study a large number of nonequilibrium systems.  For example, magnetohydrodynamic systems yield asymptotic states with either no magnetic field, or with a finite but constant (in time) magnetic field.   Researchers~\cite{Morin:IJMPB2009,Verma:PP2013} have modelled the aforementioned {\em dynamo transitions} using Eqs.~(\ref{eq:X3}, \ref{eq:X5}).  Another interesting phenomenon is the polarity reversals of the geo- and solar magnetic fields.   The solar magnetic field flips quasi-periodically every eleven years.  However, the interval between two consecutive reversals in geomagnetic field is random with the average interval being 200000 years.   Researchers~\cite{Petrelis:PRL2009} have attempted to model the dynamics of reversals using  amplitude equations that have similar forms as Eqs.~(\ref{eq:X3}, \ref{eq:X5}).

Systems with two stable configurations are referred to as bistable systems.  Bistability is  observed in lasers (e.g., excited and non-excited states of atoms)~\cite{Abraham:RPP1982}, climate (e.g., ice age and warm phase on the Earth)~\cite{Hawkins:GRL2011},  ecology (e.g., species proliferation and species extinction)~\cite{Folke:AREES2004}, and in transition to turbulence~\cite{Pomeau:PD1986,Pomeau:CRM2015}.  In addition to the critical parameter, such systems are affected by the noise of the environment.   Hence stochasticity is often invoked to model such system.  A stochastic version of the dynamical systems are referred to as  {\em stochastic bistable systems},  and they have been studied in  literature.  Haken~\cite{Haken:book:Synergetics} studied nonequilibrium phase transitions and pattern formation using stochastic versions of Eqs.~(\ref{eq:X3}, \ref{eq:X5}).   Shi {\em et al.}~\cite{Shi:PRE2016} studied stochastic bistable systems and classified the system behavior based on temporal scales and noise strength.  A related phenomenon is  a {\em stochastic resonance}~\cite{Gammaitoni:RMP1998} in which a weak signal is amplified by the  combined effects of nonlinearity and noise.  

In this paper we focus on Eq.~(\ref{eq:X5}) that exhibits stable states as $X=0$ and $X=X_{S^\pm}$ (see Sec.~\ref{sec:hysteresis}).    We analyze some of the system properties from a dynamical perspective. For example, for zero or weak noise, the system exhibits hysteresis.  However it exhibits phase coexistence when the noise strength becomes comparable to the potential barrier~\cite{Horsthemke:book}. We also demonstrate that the final state of the system depends on the initial condition. 

The organization of the paper is as follows: In Sec.~\ref{sec:hysteresis} we recapitulate the hysteresis behavior of Eq.~~(\ref{eq:X5}).  In Sec.~\ref{sec:noise_on_bistable_system}, we show how a particular choice of noise strength leads to transition from one phase to another, with a phase coexistence state in between.  In Sec.~\ref{sec:hysteresis_shrinkage} we demonstrate how a hysteresis can shrink under weak noise.  We conclude in  Sec.~\ref{sec:conclusion}.  In Appendix~\ref{sec:appA}, we derive equation of motion for the stochastic damped nonlinear oscillator, and in Appendix~\ref{sec:appB}, we show hysteresis and phase coexistence for another bistable system.

\section{Noiseless bistable system: Hysteresis}
 \label{sec:hysteresis}
 
 A popular dynamical model for  the first-order transition is 
\begin{equation}
\dot{X} = cX + X^3 - X^5, 
\label{eq:dynamical}
\end{equation}
where $X$  represents the mean field, and $c$ is a system parameter~\cite{Landau:book:StatMech,Reichl:book:StatMech,Pathria:book,Schwabl:book,Chaikin:Book}. In Appendix~\ref{sec:appA}, we relate the above equation to the damped nonlinear oscillator
\begin{equation}
\gamma \dot{X} = -X + \gamma (X^3 -   X^5) = F(X), 
\label{eq:damped oscillator}
\end{equation}
where $X,\dot{X}$ are respectively the amplitude and velocity of the oscillator, $\gamma=1/|c|$ is the frictional coefficient, and $F(X)$ is the external force.  We assume that the natural frequency of the oscillator is unity, and $X^3, X^5$ are the nonlinear terms. Here, the viscous damping is so strong that it matches with the external force; it is the quasi-static limit of the oscillator.   

\begin{figure}[htbp]
\begin{center}
\includegraphics[scale=0.8]{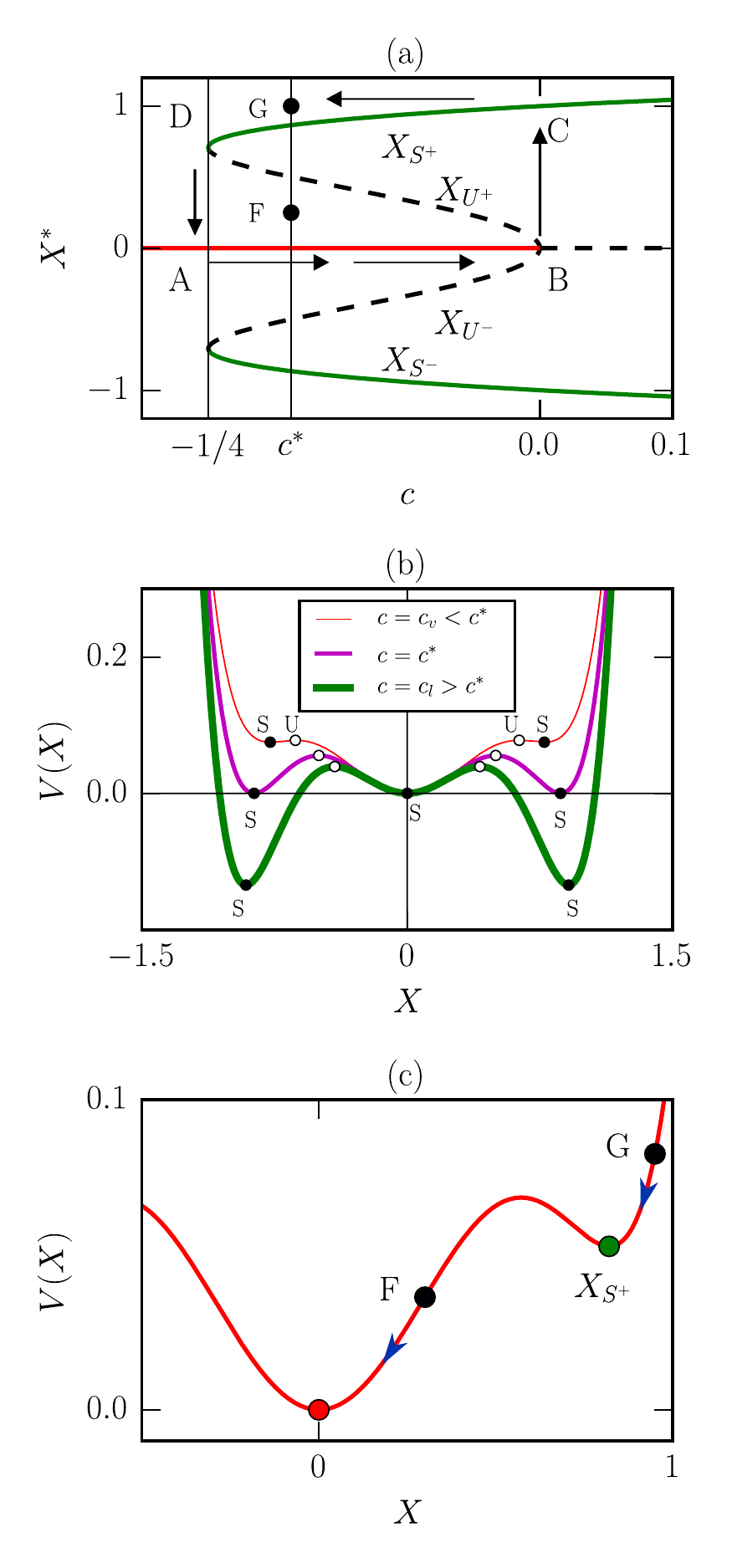}
\end{center}
\setlength{\abovecaptionskip}{0pt}
\caption{ {The fixed points $X^{*}$ of $\dot{X} = cX + X^3-X^5 = cF(X)$ and the potential $V(X)=- \int^X F(X^\prime) dX^\prime$.} (a) Stable fixed points $X=0$ and $X_{S^{\pm}}$, and unstable points  $X_{U^{\pm}}$. (b) Plots of $V(X)$ for $c=c_v$, $c^{*}$, and $c_l$.  Note that $c_v < c^*$ and $c_l > c^*$.   (c) For a parameter $c$, a system with the initial condition F reaches $X=0$, while one with G reaches $X=X_{S^{+}}$, thus exhibiting sensitivity to initial condition.}
\label{fig:pot}
\end{figure}

The steady-state solutions of  Eq.~(\ref{eq:dynamical}), obtained by setting $\dot{X}=F(X)=0$, are shown in Fig.~\ref{fig:pot}(a). The stable solutions, $X = 0$ and
\begin{equation}
X_{S^{\pm}} = \pm \sqrt{\frac{1}{2} + \sqrt{\frac{1}{4} + c} },
\label{eq:stable}
\end{equation}
are shown as red and green curves respectively, while  
\begin{equation}
X_{U^{\pm}} = \pm \sqrt{\frac{1}{2} - \sqrt{\frac{1}{4} + c} },
\label{eq:unstable}
\end{equation}
the unstable solutions, are shown as dashed black curves~\cite{Strogatz:book}.  When we relate the above dynamical system to the liquid-vapor transition,  the red curve corresponds to the vapor state, while the green curves to the symmetric liquid states.  In the parameter regime $[-1/4,0]$, the system can take one of the three stable solutions $X$, $X_{S^+}$, and $X_{S^-}$.  

The system corresponding to Eq.~(\ref{eq:dynamical}) exhibits hysteresis, as described below.  Suppose the system is at A on the $X = 0$ branch of Fig.~\ref{fig:pot}(a).  On an increase of $c$, the system remains on the $X = 0$ branch until B, after which it jumps to C. When we decrease $c$ at C, the  system follows $X_{S^+}$ branch until D. A further decreases of $c$ pushes the system to A. This phenomenon, called {\em hysteresis}~\cite{Landau:book:StatMech,Reichl:book:StatMech,Pathria:book,Schwabl:book,Strogatz:book}, is observed in many physical examples, e.g., for a ferromagnet in an external magnetic field, and in atmosphere~\cite{Hawkins:GRL2011}, optics~\cite{Abraham:RPP1982}, and ecology~\cite{Folke:AREES2004},  dynamo transition~\cite{Morin:IJMPB2009,Verma:PP2013},  etc.  It is interesting to note that the final state of the system depends on the initial condition.  A system with the initial condition at F of  Fig.~\ref{fig:pot}(c) goes to $X=0$, while one at G settles down to $X_{S^+}$.  This feature has not been explored extensively in many applications of statistical physics since time dependence ($d/dt$) is typically absent in equilibrium statistical physics.  We believe that an exploration of initial condition sensitivity in experiments could yield interesting predictions for such systems.

The potential 
\begin{equation}
V(X) = - \int^X F(X^{\prime})dX^{\prime},
\label{eq:pot}
\end{equation}
displayed in Fig.~\ref{fig:pot}(b), also helps us understand the dynamics of transition.  The potential minima corresponding to the stable solutions, marked as S in Fig.~\ref{fig:pot}(b), are separated from each other by the potential maxima corresponding to the unstable solutions, which are marked as U in Fig.~\ref{fig:pot}(b).  The three curves correspond to $c=c_v, c^*$, and $c_l$. The global minimum for $c=c_v$ and $c_l$ are at $X=0$ and $X_{S^\pm}$ respectively.  At $c=c^*$, the potential values at $X=0$ and $X_{S^\pm}$ are the same.

From the potential plot too, we demonstrate that the final state of the system depends on the initial condition.  For example, in Fig.~\ref{fig:pot}(c), a system with configuration F will settle down to $X = 0$, while configuration G settles down to $X=X_{S^+}$.  Note that the system slides to the local minima, which may not be the global minimum. Verma and Yadav~\cite{Verma:PP2013} demonstrated  the initial condition dependence of the final state  in a dynamo transition.  They performed numerical simulations of magnetohydrodynamic equations with two different initial conditions for the same parameter values, and observed that one initial condition leads to zero magnetic field, while the other one yields nonzero magnetic field.

In the next section, we introduce noise in Eq.~(\ref{eq:dynamical}) and study transition and state coexistence \, in \, a \, stochastic \, bistable system.

\section{Transition and state coexistence in stochastic bistable system}
\label{sec:noise_on_bistable_system}

\begin{figure*}[htbp]
\begin{center}
\includegraphics[scale = 0.9]{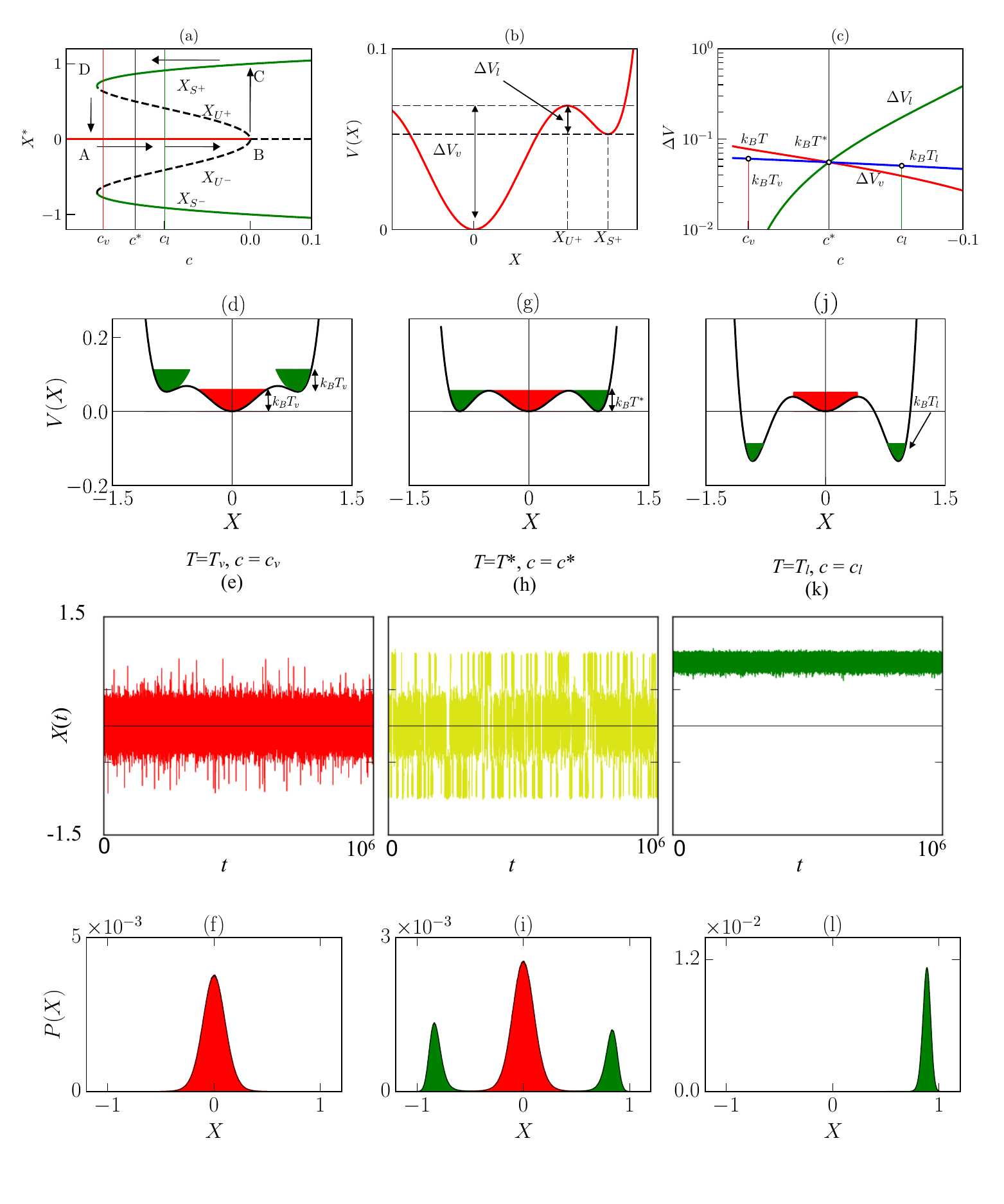}
\end{center}
\setlength{\abovecaptionskip}{0pt}
\caption{{For the transition and phase coexistence in stochastic two-state system:} (a) Fixed point $X^{*}$ vs $c$. (b) $V(X)$ vs. $X$ depicting the potential differences $\Delta V_v = V(X_{U^+}) - V(0)$ and $\Delta V_l = V(X_{U^+}) - V(X_{S^+})$. (c) Plots of  $\Delta V_v$, $\Delta V_l$, and temperature $T = T^{*}-0.1(c-c^*)$, where  $T^*$ is the critical temperature. For the vapour state, $T = T_v$ or $c = c_v$: (d) $V(X)$ vs. $X$ with the shaded region exhibiting thermal energy $k_B T_v$ with $\Delta V_l<k_B T_v<\Delta V_v$; (e) The time series $X(t)$ vs. $t$; (f) Probability distribution function PDF $P(X)$.  At the transition temperature, $T = T^*$ or $c = c^*$: (g,h,i) Plots of $V(X)$ vs. $X$, $X(t)$ vs. $t$, and $P(X)$.   For the liquid state, $T = T_l$ or $c = c_l$: (j,k,l) Plots of $V(X)$ vs. $X$, $X(t)$ vs. $t$, and $P(X)$.}
\label{fig:time}
\end{figure*}

The  system  described by Eq.~(\ref{eq:dynamical}) is noiseless, hence it can be treated to be at zero temperature.   A natural question is  whether the above system could exhibit phase coexistence, as in solid-liquid and liquid-vapor phase transitions.  We show below how an introduction of noise can  yield such features.  With noise, the dynamical Eq.~(\ref{eq:dynamical}) is modified to  
\begin{equation}
\dot{X} = cX + X^3 - X^5 + \eta(t) 
\label{eq:dynamical_noise}
\end{equation}
with  
\begin{equation}
\langle \eta(t) \eta(t^\prime) \rangle= 2 |c| k_B T\delta(t-t^\prime),
\label{eq:noise}
\end{equation}
 where $T$ and $k_B T$ are interpreted as the temperature and thermal energy  respectively~\cite{Reichl:book:StatMech,Chaikin:Book} (see Appendix~\ref{sec:appA}). Here $k_B$ is the Boltzmann constant.  In the following discussion, we relate the $X = 0$ state to the gaseous phase, while $X=X_{S^+}$ to the liquid phase.  The thermal energy works against the potential barriers: 
\begin{equation}
\Delta V_v = V(X_{U^+}) - V(0)
\label{eq:pot_vapour}
\end{equation}
for the gaseous state and the unstable barrier, and
\begin{equation}
\Delta V_l = V(X_{U^+}) - V(X_{S^+})
\label{eq:pot_liquid}
\end{equation}
for the liquid state and the unstable barrier.   These barriers are shown in Fig.~\ref{fig:time}(b).

For modeling phase transitions under heat bath, it is customary to take $c \propto (T-T^*)$, where $T^*$ is the transition temperature~\cite{Landau:book:StatMech,Reichl:book:StatMech,Pathria:book,Schwabl:book,Strogatz:book}.  Hence the system parameter $c$  is a function of temperature.  We invert the above relation to 
\begin{equation}
T = T^* - 0.1(c-c^*)
\label{eq:temp}
\end{equation}
with $c = c^*$ at $T = T^*$, as shown in Fig.~\ref{fig:time}(c).  Note that the temperature decreases as $c$ increases or $\gamma$ decreases. The temperature can take the system from one potential minimum to another minimum if the thermal energy is sufficiently large to climb the potential barrier corresponding to the unstable configuration.  Such jumps are not possible without noise.  

Equation~(\ref{eq:dynamical_noise}) is a stochastic differential equation (SDE), which can be analyzed analytically~\cite{Haken:book:Synergetics} or numerically. In this paper, we solve this equation using numerical simulation.  We solve Eq.~(\ref{eq:dynamical_noise}) using second-order Runge-Kutta (RK2) method prescribed by Roberts~\cite{Roberts:book:complex}.  From Eq.~(\ref{eq:noise}), we can deduce the form of $\eta$  as 
\begin{equation}
\eta = \sqrt{\frac{2|c| k_B T}{\Delta t}}\xi,
\label{eq:noise_form}
\end{equation}
where $\xi$ is the Gaussian noise with zero mean.  We divide the time domain into $n$ equal intervals with $(n+1)$ data points including the endpoints.  Here the step size $\Delta t = t_{n+1} - t_n$. A time step of Eq.~(\ref{eq:dynamical_noise}) involves
\begin{equation}
X_{n+1} = X_n + \frac{k_1 + k_2}{2},
\end{equation}
where~\cite{Roberts:book:complex}
\begin{eqnarray}
k_1 & = & \Delta t [f(X_n, t_n) + \eta_n + S_n], \\
k_2 & = & \Delta t [f(X_n + k_1, t_n + \Delta t) + \eta_n - S_n].
\end{eqnarray}
Here $S_n = \pm 1$ with each alternative chosen with equal probability of $1/2$, and $\eta_n$ is a random variable with a uniform distribution given by Eq.~(\ref{eq:noise_form}). We performed our simulation for $2$ million time steps with a fixed $\Delta t = 0.1$.  We report the steady-state statistics of the system using the later half of the time series  so as to discard the transients.

In the following discussion, we will examine the state of the system at three temperatures---$T_v$, $T_l$, and $T^*$ shown in Fig.~\ref{fig:time}(c); the corresponding values of $c$ for these temperatures are $c_v$, $c_l$, and $c^*$  respectively (see Fig.~\ref{fig:time}(a)). We will show below that they correspond to the vapor state, the liquid state, and phase-coexistence during a liquid-vapor transition.

At $T = T_v$ or $c = c_v$, thermal energy $T_v$ has property that $\Delta V_l < k_B T_v < \Delta V_v$, as shown in Fig.~\ref{fig:time}(c,d). Hence the thermal energy due to temperature is strong enough to overcome $\Delta V_l$  barrier. Therefore the system escapes from the liquid state, and remains in the vapor state.  We solve the Eq.~(\ref{eq:dynamical_noise}) numerically with random noise corresponding to $T = T_v$ and obtain the time series $X(t)$ and the probability distribution of $X$, both of which are plotted in Figs.~\ref{fig:time}(e,f) respectively.  Clearly, $X(t)$ fluctuates around $X = 0$, the global minimum, and the rms value of the fluctuations is proportional to the temperature.  There is, however, a small probability for the system to jump from the global minimum to local minima  with the rate given by Kramer's rule~\cite{Reichl:book:StatMech}.

At $T = T_l$ or $c = c_l$, $\Delta V_v < k_B T_l < \Delta V_l$,  as shown in Fig.~\ref{fig:time}(j). Following similar arguments as described above, we conclude that the temperature will take away the system from the vapor state  ($X = 0$) and push it to the liquid state ($X=X_{S^+}$ or $X_{S^-}$). The time series of $X$ and its probability distribution, shown in Figs.~\ref{fig:time}(k,l) also indicate that the system gets locked into one of the two liquid states. 

Lastly, at $T = T^*$ or $c=c^*= -3/16$, $ k_B T^* = \Delta V_l = \Delta V_v$,  as shown in Fig.~\ref{fig:time}(g).  Here, the temperature makes the system fluctuate between the liquid and vapor states as indicated    in    the      time   series   $X(t)$ and   its   probability distribution (shown in Figs.~\ref{fig:time}(h,i)).  This is the phase-coexistence phenomena. In statistical mechanics and thermodynamics, the condition $ k_B T^* = \Delta V_l = \Delta V_v$ corresponds to equating the free energies of the two phases at the transition, as in Maxwell's construction~\cite{Landau:book:StatMech,Reichl:book:StatMech,Pathria:book,Schwabl:book}.

Thus, for the potential profile of Fig.~\ref{fig:time}(b), at a given temperature, the stochastic system moves to the global minimum at that temperature.  We observe that the two states coexist at $T=T^*$. Thus our stochastic model approximately mimics the thermodynamic mean-field behavior.  Our calculation also reveals that the critical temperature $T^*$ is related to the barrier height between the potentials of unstable and stable configurations at the transition.  It is a useful prediction, and it requires information about the unstable configuration. 

It is interesting to relate the above phenomena to thermodynamics.  It is tempting to relate the transition $T^*$ to the latent heat $L$ between the two phases, i.e. $k_B T^* = L$.  But this is not the case. A careful investigation shows that the free energy   is typically reported for a stable configuration (for example, for liquid/vapor/solid phases).  Equating the free energies of the two phases that are participating in phase transition, we obtain 
\be
F =E_1 -T^* S_1 =E_2 -T^* S_2,
\ee
where $E_i$ and $S_i$ are the internal energies and the entropies of the two states. Hence 
\be
\Delta E = T^* \Delta S = L,
\ee
where $ \Delta S$ is the change in entropy between the two phases. Therefore, 
 \be
 \frac{L}{R T^*} = \frac{ \Delta S}{R},
 \ee
 where $R$ is the gas constant, and $ \Delta S$ is units of Joule/ (mole Kelvin).  According to Trouton's rule, $L/RT^* \approx 8$ to 15 for liquid-vapor transition, and approximately 1 to 3 for solid-liquid or melting transition for a wide range of materials at normal pressure and temperature~\cite{Bernstein:AJP1987}.  Thus we show that the $L \ne R T^*$.  Rather, $T^*$ is determined from the potential barrier height of the stable phase and unstable configuration.  To best of our knowledge, entropy computation of the unstable configuration between the two stable phase has not been performed.  For example, for the liquid-vapor transition, the system should be making a transition from liquid to vapor after climbing the potential barrier corresponding to the unstable configuration.  It will be interesting to study this configuration in Monte-Carlo simulations of phase transitions.   

In the present paper we relate the parameter $c$ to the temperature $T$ in order to model thermally-induced transitions. Note however that different temperature profiles will exhibit different behavior. Also, $c$ and $T$ could be varied independently that would give flexibility in modelling different complex material.  

The features discussed in this section is generic, and it is observed in other bistable systems.   In Appendix~\ref{sec:appB} we show that the system 
\be
\dot{X} = -\lambda + 3 X - X^3 
\ee
exhibits similar features as the system corresponding to Eq.~(\ref{eq:X5}).  Recently,   Shi~{\em et al.}~\cite{Shi:PRE2016} studied the noisy version of above system and related the noise strengths to state transition, basin transition, and distribution transition.  In our paper we provide simpler dynamical interpretations. 

\begin{figure}[htbp]
\begin{center}
\includegraphics[scale=0.8]{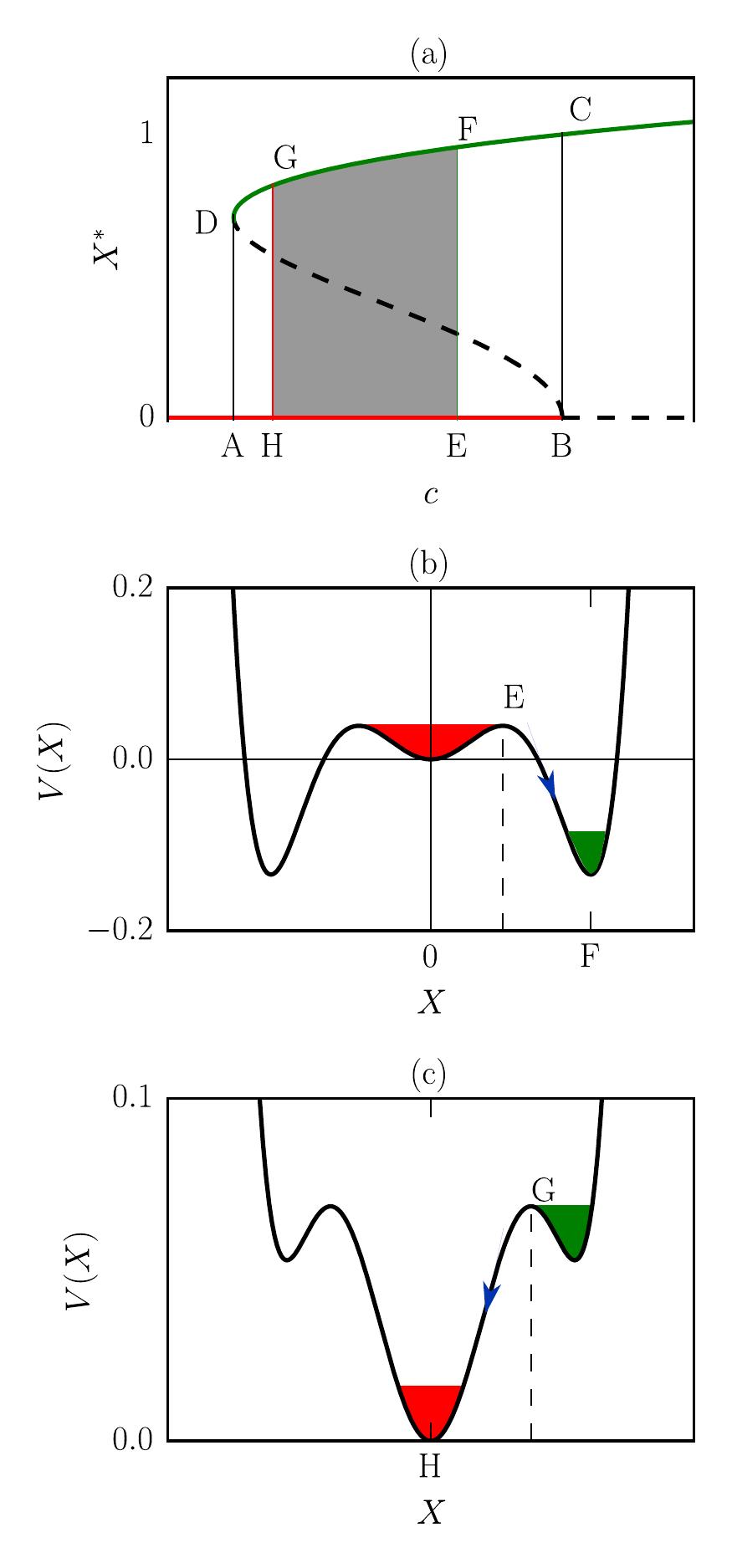}
\end{center}
\setlength{\abovecaptionskip}{0pt}
\caption{{Shrinkage of hysteresis with a weak noise:} (a) ABCD is the hysteresis without noise, but EFGH is one with noise. (b) In $V(X)$ vs. $X$ plot, the noise induces transition from E to F before B.  (c) Due to the noise, the system transitions from G to H before D.}
\label{fig:hyst}
\end{figure}

Stochastic bistable systems also exhibit other interesting features, e.g., hysteresis shrinkage, which will be discussed in the next section.

\section{Hysteresis shrinkage due to noise}
\label{sec:hysteresis_shrinkage}

The hysteresis of Fig.~\ref{fig:pot} has no noise.  In stochastic bistable systems, the hysteresis shrinks  under the  introduction of noise or temperature.  As  \, discussed  \, earlier,  \, the \, zero-temperature hysteresis is given by ABCD loop of Fig.~\ref{fig:hyst}(a).  However, an introduction of weak noise can yield a transition from $X = 0$ to $X = X_{S^+}$  at E of Fig.~\ref{fig:hyst}(a) rather than at B, and from $X = X_{S^+}$ to $X = 0$ at G rather than at D.  The corresponding potential functions are shown in Fig.~\ref{fig:hyst}(b,c) respectively. Thus, in the presence of noise, the width of the hysteresis has shrunk from ABCD to HEFG of Fig.~\ref{fig:hyst}(a).

Such hysteresis shrinkage have been reported in other areas of science.  For example, Guttal and Jayaprakash~\cite{Guttal:EM2007} showed that region of bistability in environmental system  (characterised by parameters such as nutrient input and rainfall) are reduced under the introduction of weak noise.  Similarly, Gopalkrishnan and Sujith~\cite{Gopalakrishnan:JFM2015} observed shrinkage of the hysteresis loop in Rijke tube under the introduction of weak noise.   

The above analysis shows that a parametric study of the noise strength and $c(T)$  yields interesting insights into these systems.  

\section{Conclusion}

\label{sec:conclusion}
In this paper we present  effects of noise in bistable systems. For zero noise or a small amount of noise, the system shows hysteresis, but the hysteresis width decreases with the increase of noise. By appropriate modelling of the noise strength with system parameters, we can model the transition in the system from one state to another, along with a phase coexistence at the critical temperature.  Thus a single stochastic system can exhibit hysteresis and phase coexistence.   Such features could be useful for studying  systems in optics, ecology, economics, etc.

We thank Stephan Fauve, Maurice Rossi, Amit Dutta, Daan Frenkel, R. Sujith, and Arul Lakshminarayam for valuable suggestions. This work was supported by the research grants SERB/F/3279 from Science and Engineering Research Board, India, and PLANEX/PHY/2015239 from Indian Space Research Organisation, India.


\section*{Appendix A: Stochastic nonlinear damped oscillator under extreme dissipation}
\label{sec:appA}

The Langevin's equation with an external constant forcing $F$ and noise $\zeta(t)$ is
\be
 m \dot{v} + \gamma v = F(X) + \zeta(t).
\ee
The above equation describes the motion of a particle of mass $m$ experiencing frictional force $-\gamma v$, external force $F$, and random force $\zeta(t)$.  The position and velocity of the particle $X$ and $v$ respectively.  The solution of the above equation is~\cite{Reichl:book:StatMech}
\bea
 v(t) & = & v_0 \exp(-\gamma t/m) 
 +  \bigg[ \frac{1}{m} \exp(-\gamma t/m) \nonumber \\ &\times & \int^t [F+\zeta(t')] \exp(\gamma t'/m) dt' \bigg].
\eea
For extreme dissipation, the transient, $\exp(-\gamma t/m) \rightarrow 0$, hence
\be
v = \dot{X} = \frac{1}{\gamma} ( F + \zeta(t)), \label{eq:dot_v}
\ee
where $X$ is the position of the particle.  For Langevin's equation~\cite{Reichl:book:StatMech}
\be
\langle \zeta(t) \zeta(t') \rangle = 2 \gamma k_B T  \delta(t-t'), 
\label{eq:zeta_zeta}
\ee
where $T$ is the temperature, and $k_B$ is the Boltzmann's constant.  

We choose the external force as
\begin{equation}
F(X) = -X + \gamma (X^3 -   X^5), 
\label{eq:damped oscillator}
\end{equation}
and $\zeta = \gamma \eta$, then 
\begin{equation}
\dot{X} = -c X + \gamma (X^3 -   X^5) + \eta
\label{eq:damped oscillator_appendix}
\end{equation}
with $\gamma = 1/|c|$. 
Thus, the system corresponds to a nonlinear stochastic oscillator with natural frequency $\omega_0=1$. Also, using Eq.~(\ref{eq:zeta_zeta}), we obtain
\be
\langle \eta (t) \eta(t') \rangle = \frac{1}{\gamma^2} \langle \zeta(t) \zeta(t') \rangle  =2 |c|k_B T  \delta(t-t').
\ee
We employ the above equation for the simulation of the stochastic bistable system.

\section*{Appendix B: Hysteresis and phase coexistence in $\dot{X}=-\lambda +3X-X^{3}+\eta$}
\label{sec:appB}

In this section, we  study another noisy bistable system and show that this system too exhibits hysteresis and phase coexistence as  discussed in  Sections \ref{sec:hysteresis} and \ref{sec:noise_on_bistable_system}.  This example is often used to describe first-order phase transition~\cite{Chaikin:Book}.

\begin{figure}[htbp]
\begin{center}
\includegraphics[scale=0.6]{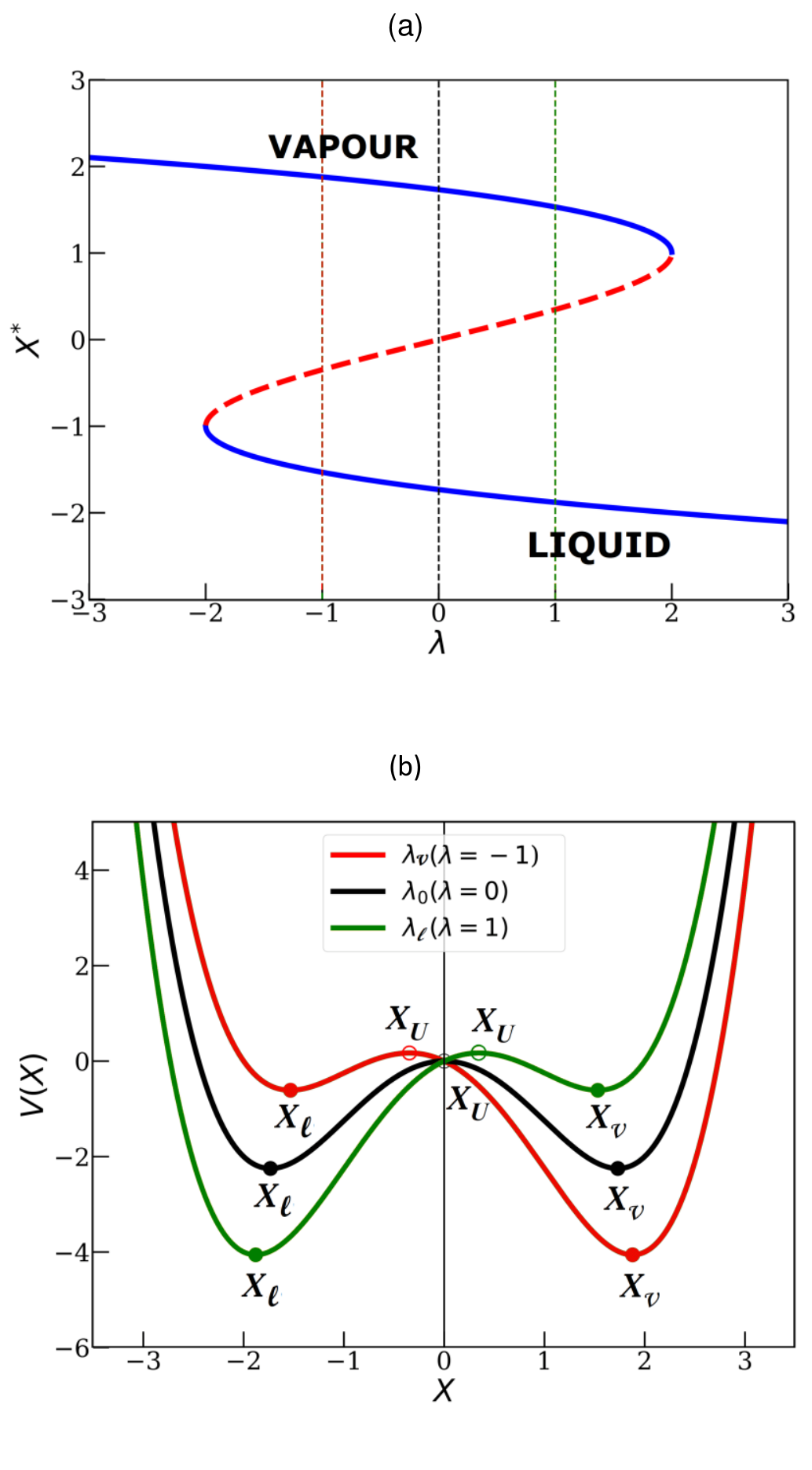}
\end{center}
\setlength{\abovecaptionskip}{0pt}
\caption{(a) Bifurcation diagram corresponding to Eq.~(\ref{eq:Xdot_sys2}); the upper and lower branches represent  vapour and liquid states respectively. (b) Plots of $V(X)$ vs.~$X$ for $\lambda = \lambda_{l} =1$ (green curve, liquid state), $\lambda = \lambda_{*}=0$ (black curve, phase coexistence), and $\lambda = \lambda_{v}= -1$ (red curve, vapor state).}
\label{fig:bifurcation}
\end{figure}

First, we describe the new bistable system.  The potential function of the system is  
 \begin{equation}
V(X)=\lambda X-\frac{3}{2}X^{2}+\frac{1}{4}X^{4},
\label{eq:pot_sys2}
\end{equation}
where $\lambda$ is the control parameter, and the corresponding stochastic differential equation is \begin{equation}
\dot{X}=-\lambda +3X-X^{3}+\eta \left ( t \right ),
\label{eq:Xdot_sys2}
\end{equation}
where $\eta \left ( t \right )$ is the noise strength with
\begin{equation}
\label{eq:T_sys2}
\left \langle \eta \left ( t \right )\eta \left ( {t}' \right ) \right \rangle = 6k_{B}T\delta \left ( t-{t}' \right ), 
\end{equation}
where $k_{B}T$  represents the thermal energy\cite{Reichl:book:StatMech}.

First we consider the noiseless version of Eq.~(\ref{fig:bifurcation}).  In Fig.~\ref{fig:bifurcation}(a) we plot the bifurcation diagram in which the the $x$-$y$ axes represent $\lambda$ and the fixed point $X^*$ respectively.  From the figure we deduce that the system has three fixed points---$X_l, X_v$ (stable ones) and $X_u$ (unstable).  Here $l$  and $v$ represent the liquid and vapour states.  These  interpretations become apparent when we consider the stochastic equation.

\begin{figure}[htbp]
\begin{center}
\includegraphics[scale=0.6]{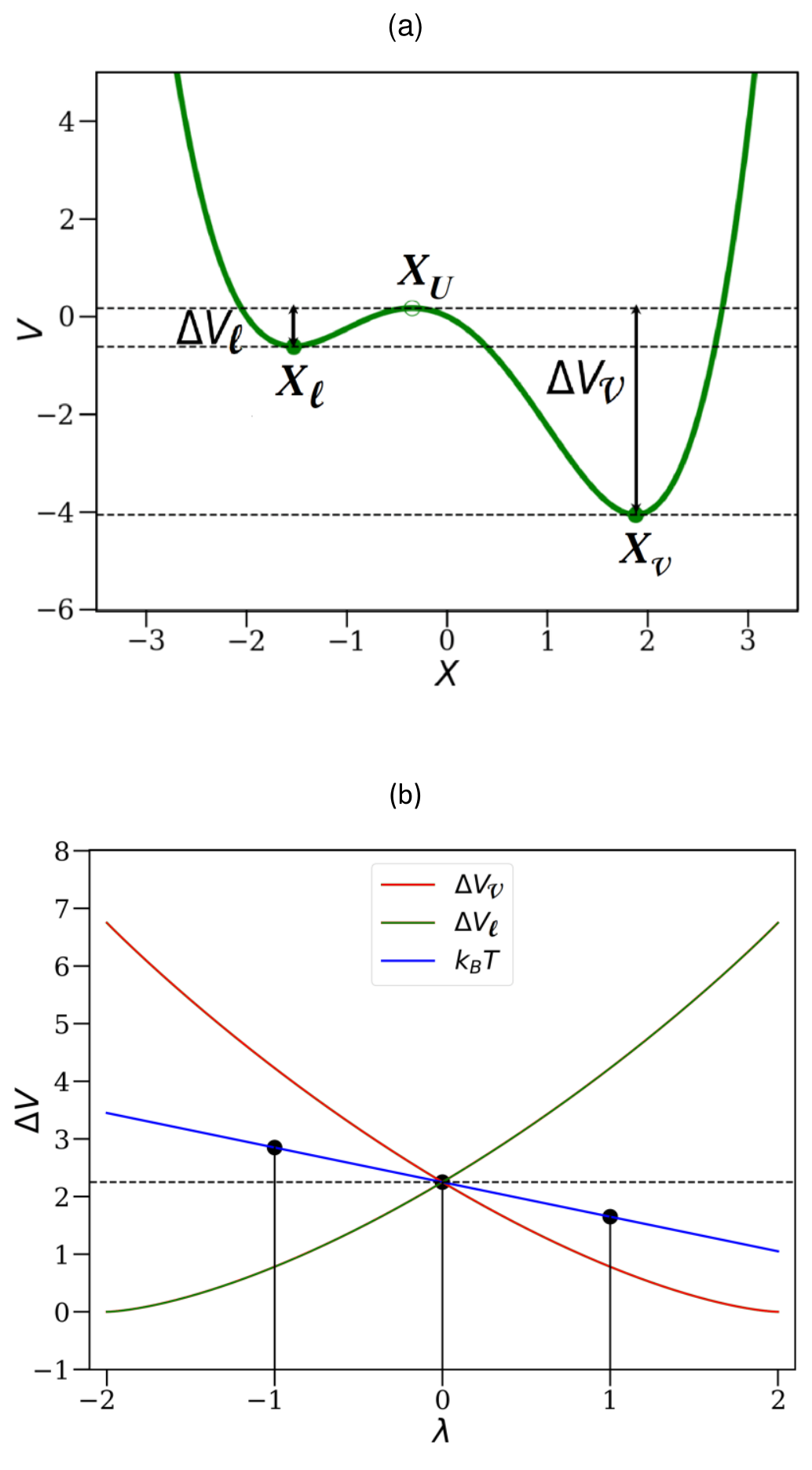}
\end{center}
\setlength{\abovecaptionskip}{0pt}
\caption{(a) Plot of $V(X)$ vs.~$X$ illustrating $\Delta V_{l}$ and $\Delta V_{v}$. (b) Plots of $\Delta V_{l}$, $\Delta V_{v}$, and $T=T^{*}-0.6(\lambda-\lambda^{*})$ vs.~$\lambda$.}
\label{fig:deltaV}
\end{figure}

In Fig.~\ref{fig:bifurcation}(b) we plot potential function for three cases---$\lambda=\pm 1, 0$.  For $\lambda=-1$, the global minimum occurs with $X^*>0$ (vapour state).  Hence, we label $\lambda=-1$ as $\lambda_v$.  On the contrary, the global minimum for $\lambda=1$ has $X^*<0$ (liquid state), hence we label $\lambda=1$ as $\lambda_l$.  For $\lambda=0$, $V(X_l) = V(X_v)$, hence $\lambda=0$ corresponds to phase coexistence, and we denote it by $\lambda_*$. Note that the maxima of the potential corresponds to an unstable phase.  Since the system has two potential minima and a potential maximum, it has two potential barriers as shown in Fig.~\ref{fig:deltaV}(a):
\begin{equation}
\label{eq:pot_sys2}
    \begin{split}
        \Delta V_{l}=V(X_{U})-V(X_{l}),\\
    \Delta V_{v}=V(X_{U})-V(X_{v}).
    \end{split}
\end{equation}

\begin{figure*}[htbp]
\begin{center}
\includegraphics[scale = 0.6]{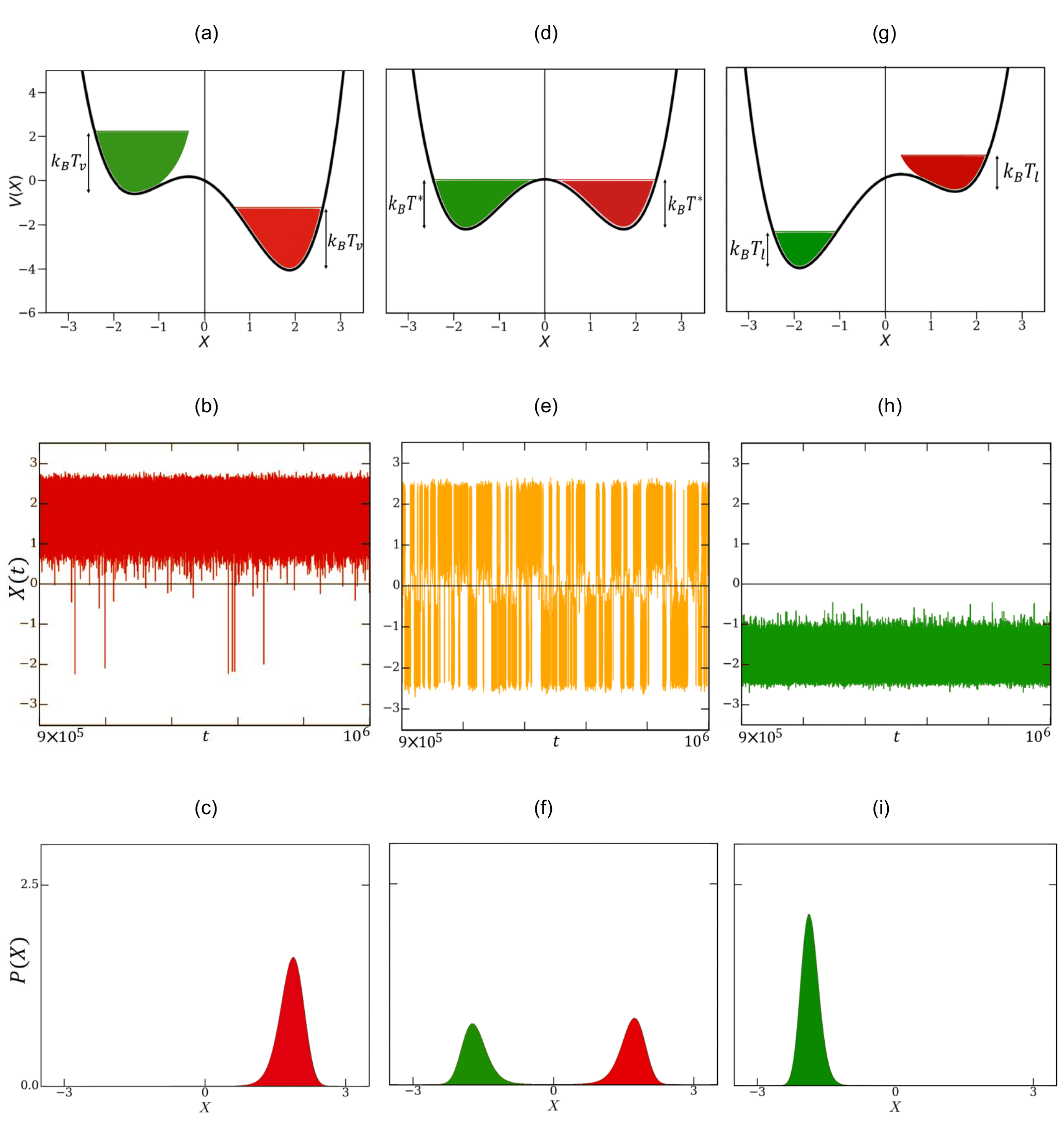}
\end{center}
\setlength{\abovecaptionskip}{0pt}
\caption{Refer to $\dot{X}=-\lambda +3X-X^{3}+\eta \left ( t \right )$ [Eq.~(\ref{eq:Xdot_sys2})]. (a,b,c) For the vapor state with $T=T_v$, plots of potential $V(X)$, time series $X(t)$, and PDF of $X$.  (d,e,f) Corresponding plots for the phase-coexistence.  (g,h,i) Corresponding plots for the liquid state. The notation is same as Fig.~\ref{fig:time}. Here red and green colours refer to the vapor and liquid states respectively.}
\label{fig:time2}
\end{figure*}

Now we introduce noise in the system and solve using the same procedure as in Sec.~\ref{sec:noise_on_bistable_system}.  We employ temperature given by 
\begin{equation}
\label{eq:temperature_sys2}
    T=T^{*}-0.6(\lambda-\lambda^{*}),
\end{equation}
which is illustrated graphically in  Fig.~\ref{fig:deltaV}(b).  In the figure we label the temperature corresponding to $\lambda_{l,v,*}$ as $T_{l,v,*}$.  Note that $T_*$ corresponds to phase coexistence state, while $T < T_*$ to liquid state, and $T>T_*$ to the vapour state.  For a given temperature, we also solve Eq.~(\ref{eq:Xdot_sys2}) numerically using the method described in Sec.~\ref{sec:noise_on_bistable_system}.  In Fig.~\ref{fig:time2} we exhibit the potential $V(X)$, time series $X(t)$, and PDF $P(X)$ for the liquid and vapor states, as well as for the coexistence regime.

For $\lambda=-1$ and $T=T_v$, as shown in Fig.~\ref{fig:deltaV}(b) and Fig.~\ref{fig:time2}(a), the thermal energy $k_B T_v$ exceeds the potential barrier $\Delta V_l$, but it is less than $\Delta V_v$. Hence the system fluctuates around the fixed point $X_v$, the vapour state. This is evident from the time series $X(t)$ and PDF $P(X)$ shown in Fig.~\ref{fig:time2}(b,c) respectively.  Note that occasionally, the system crosses $X=0$ barrier.

For $\lambda=1$ and  $T=T_l$, as shown in Fig.~\ref{fig:deltaV}(b) and Fig.~\ref{fig:time2}(g), $\Delta V_v < k_B T_v < \Delta V_l$.  Hence, the system tends to oscillate around $X=X_l$, the liquid state. See Fig.~\ref{fig:time2}(h,i) for illustrations of the time series $X(t)$ and PDF $P(X)$. For this case, the fluctuations are lower than that for  $T=T_v$ due to the lower temperature.  For $T=T_*$, $\Delta V_v = k_B T_v = \Delta V_l$, hence the system fluctuates around both the minima, and it corresponds to the phase coexistence.  

Thus, we show that Eq.~(\ref{eq:Xdot_sys2}) exhibits two phases and phase coexistence depending on the temperature.   The aforementioned example and that of Sec.~\ref{sec:noise_on_bistable_system} demonstrate that these features are common among the noisy bistable systems.




\end{document}